\documentstyle[12pt]{article}

\advance\hoffset by -7mm
\makeatletter
\@addtoreset{equation}{section}
\makeatother

\renewcommand{\title}[1]{\null\vspace{25mm}

\noindent{\Large{\bf #1}}\vspace{10mm}

\noindent {\large By }}
\newcommand{\authors}[1]{\noindent{\large #1}\vspace{3mm}

}
\newcommand{\address}[1]{\noindent #1\vspace{5mm}

}
\renewcommand{\abstract}[1]{\vspace{19mm}

\noindent{\small{\em Abstract.} #1}\vspace{2mm}

}

\newcommand{\theomod}[1]{\noindent{\bf Theorem #1}{\hspace{0em}}}
\newcommand{\corollary}[1]{\noindent{\bf Corollary #1}{\hspace{0em}}}
\newcommand{\mlemma}{\noindent{\bf Main Lemma.}{\hspace{0em}}}
\newcommand{\Ad}{\mbox{\rm Ad}}
\newcommand{\lbrac}[2]{[#1,#2]}
\renewcommand{\l}{\lambda}

\begin{document}

\begin{flushright}
Z\"urich University Preprint\\
ZU-TH 29/96\\
gr-qc/9610024\\
\end{flushright}
\vspace{-9 ex}

\title{On Symmetric Gauge Fields for arbitrary\protect\\ Gauge and
Symmetry Groups}
\authors{Othmar Brodbeck
}
\address{Institute for Theoretical Physics, University of
Z\"urich,\\ Winter\-thur\-er\-stras\-se\nobreak\ 190, 8057 Z\"urich,
Switzerland
}
\abstract{
A classification of the possible symmetric principal bundles with a
compact gauge group, a compact symmetry group and a base manifold which
is regularly foliated by the orbits of the symmetry group is derived.
A generalization of Wang's theorem (classifying the invariant
connections) is proven and local expressions for the gauge potential
of an invariant connection are given.
}
\section{Introduction}
A basic motivation to consider symmetric gauge fields is the reduction
of the classical degrees of freedom of a gauge field theory. However,
imposing a symmetry for gauge fields leads to a surprise similar to the
``Kaluza-Klein miracle'': The Yang-Mills action for symmetric fields is
a Yang-Mills-Higgs action in a lower dimension. This ``dimensional
reduction'' has been studied by several authors and is used, for
instance, to construct generalized Kaluza-Klein theories with an
additional Yang-Mills field and a homogeneous space as internal space.
(For a review and key references see, e.g., \cite{kubyshin}.)

One approach for a systematic discussion of symmetric gauge fields is
the following one. First, one constructs those principal bundles which
admit an action of the symmetry group (by automorphisms), such that the
induced action on the base agrees with a given one. Then, one
classifies the invariant connections in these bundles and, finally, one
derives local expressions for the gauge potentials of an invariant
connection. The resulting fields are then, by con\-struc\-tion,
invariant up to a gauge transformation. \hbox{(For an alternative
approach see \cite{manton}.)}

On the basis of previous works \cite{harnad}, we have carried out these
steps for arbitrary gauge und symmetry groups. Our results close a gap
in the literature for the bundle classification and simplify the last
two steps. Below we give a brief description of our investigations.
(Details will be published elsewhere; for partial results see {Appendix
I of \cite{OB1}.)}
\section{Symmetric bundles with a regularly foliated base}
For a symmetric principal bundle $P(M,G)$ with a compact gauge group
$G$ and a compact symmetry group $K$, we first consider the induced
$K$-action on the base manifold $M$. Since the symmetry group is
compact there exists an open and dense submanifold $M_{(H)}\subset M$
(the principal orbit bundle) which, at least locally, is regularly
foliated by the orbits of $K$. That is, locally, $M_{(H)}\approx
M_{(H)}/K\times K/H$ for a (compact) subgroup $H\subset K$
\cite{dieck}. With a minor loss of generality, we may assume,
therefore, that the base has the form
\[
M=\tilde M\times K/H\;,
\]
where $\tilde M$ is a connected manifold,
and that the $K$-action on $M$ agrees with the canonical action of $K$
on the homogeneous space $K/H$.

\theomod{1.}
{\em
With the assumptions above, a $K$-symmetric principal bundle $P(M,G)$
is classified by
a group homomorphism $\l\colon H\to G$ (out of a complete set of
non-conjugate homomorphisms) and
a principal bundle $\tilde Q(\tilde M,Z)$, where $Z$ is the centralizer
of the subgroup $\l(H)\subset G$.
}

We wish to emphasize that this theorem represents a ``global'' result
and that for a proof, no further assumptions are required.

The classifying bundle $\tilde Q$ is, as expected, a subbundle of
$P|_{\tilde M}$, the portion of $P$ over the submani\-fold $\tilde
M\cong \tilde M\times\{eH\}$. To show this, we first observe that
$\tilde M\times\{eH\}$ is a fixed point set of the subgroup $H\subset
K$. Thus, {each fiber} of $P|_{\tilde M}$ is maped onto itself by the
action of $H$. Next, let us use this property to introduce the map
\[
\mu: P|_{\tilde M}\times H\to G\;,\qquad (p,h)\mapsto\mu_{p}(h)\;,
\]
where $\mu_{p}(h)$ is defined by
\[
h\cdot p\;=\; p\cdot\mu_{p}(h)\;\;.
\]
Then, for each $p\in P|_{\tilde M}$, the restriction $\mu_p\colon H\to
G$ is a group homomorphism and for points in the same fibre of
$P|_{\tilde M}$, the corresponding homomorphisms belong to the same
conjugacy class. The following lemma now completes the construction.

\mlemma{}
{\em Let $p_0$ be an arbitrary point of $P|_{\tilde M}$. Let
$\l=\mu_{p_0}$ be the homomorphism corresponding to $p_0$ and let $Z$
be the centralizer of the subgroup $\l(H)\subset G$. Then
\[
\tilde Q(\tilde M,Z)=\Bigl\{\;\,p\in {P|_{\tilde M}}\;\mid \;
\mu_{p}=\l\;\,\Bigr\}
\]
is a reduced bundle of $P|_{\tilde M}$ with structure group $Z$.}

Conversely, a symmetric bundle $P$ can be recovered from the
classifying pair
$(\l,\tilde Q)$ by means of a standard construction. More precisely,
$P$ is ``equivalent'' to a bundle which is associated to the product
bundle
\[
P'(M,G')=\tilde Q(\tilde M,Z)\times K(K/H,H)
\]
and has $G$ as typical fiber. For later use, we give some details of
the construction. The product bundle $P'$ is, by definition, a
principal bundle over $M=\tilde M\times K/H$ with structure group
$G'=Z\times H$. Moreover, since the second factor is $K$-symmetric, the
same is true for $P'$. Next, recall that $Z$ is the centralizer of the
subgroup $\l(H)\subset G$. Thus, the homomorphism $\l\colon H\to G$
naturally extends to a homomorphism from the structure group $G'$ of
$P'$ to the structure group $G$ of $P$:
\[
\rho:G'=Z\times H\,\longrightarrow \, G\;,\qquad
(z,h)\,\longmapsto\, z\,\l(h)\;.
\]
Now, let $G'$ act on $G$ by
\[
G'\times G\,\longrightarrow \, G\;,\qquad
g'\,\longmapsto\, \rho(g')\,g\;.
\]
{Then the associated bundle $P'\times_{G'}G$ is a $K$-symmetric
$G$-principal bundle over $M$ which is equivalent to the given bundle
$P$.} (By an equivalence we mean a $K$- and $G$-equivariant bundle
isomorphism which induces the identity of the base.)
\section{Invariant connections and symmetric\protect\\ gauge
potentials}
\theomod{2.\ }(Generalized Wang theorem)
{\em
Let $P$ be a K-symmetric principal bundle classified by $(\l,\tilde Q)$
and let $\omega$ be a connection in $P$ which is invariant under the
action of $K$.
Then $\omega$ is classified by
a connection $\tilde \omega$ in $\tilde Q$ and
a scalar field $\tilde \phi$ over $\tilde Q$ with values in the linear
subspace of $LG\otimes LH\!_{\perp}{\vphantom{\big|}}^{\!\!\ast}$
defined by
\[
\Ad(\l(h))\circ\tilde \phi=\tilde \phi\circ\Ad(h)
\]
for all $h\in H$. Here, $LG$ denotes the Lie algebra of G and
$LH\!_{\perp}$ is the complement of $LH\subset LK$ with respect to an
invariant scalar product.
}

The target space of the scalar field $\tilde \phi$ is
the fixed point set of a real representation of the
isotropy group $H$. This offers the possibility for a group theoretical
discussion of the ``constraint equation'' for $\tilde \phi$. Some
technical subtleties arise, however, since the relevant representation
is real.

A transparent proof of this theorem is obtained when the structure of
the symmetric
bundle $P$ is used. To give the basic idea, let us first recall that
$P$ is equivalent to the bundle $\pi\colon P'\times_{G'}G\mapsto M$
which is associated to the principal bundle $\psi\colon P'\to M$.
Hence, the diagram
\[{
\def\normalbaselines{\baselineskip20pt
		     \lineskip3pt \lineskiplimit3pt }
\def\u_mapright#1{\smash{
		  \mathop{\longrightarrow}\limits^{\displaystyle #1}}}
\def\d_mapright#1{\smash{
		  \mathop{\longrightarrow}\limits_{\displaystyle #1}}}
\def\lmapdown#1{\bigg\downarrow
	     \llap{$\vcenter{\hbox{$#1$}}$}}
\def\rmapdown#1{\bigg\downarrow
	     \rlap{$\vcenter{\hbox{$#1$}}$}}
\matrix{
P'\times G & \u_mapright{\Psi} & P'\times_{G'}G
\rlap{}\cr
\lmapdown{\mbox{\small Pr}_1\;\;} & & \rmapdown{\pi}\cr
P'  & \u_mapright{\psi} & M\cr}
}
\]
\noindent
where
$
\Psi\colon(p,g)\mapsto\lbrac{p}{g}
$
is the canonical projection, is commutative. On the one hand, $\Psi$ is
a
$K$-equivariant bundle map between principal $G$-bundles. Additionally,
$\Psi$ is the  projection of a principal bundle with structure group
$G'$. Now, if $\omega$ is an invariant connection in $P\cong
P'\times_{G'}G$ then, due to the first property of $\Psi$, the
pull-back $\Psi^\ast\omega$ is an invariant connection in the trivial
bundle $\mbox{Pr}_1\colon P'\times G\to P'$. From this we obtain that
$\Psi^\ast\omega$ is determined by a $K$-invariant one-form over
$P'=\tilde Q\times K$. Next, taking advantage of the second property of
$\Psi$, we consider $\Psi^\ast\omega$ as the pull-back of a one-form by
the canonical projection of a principal bundle. This time it follows
that $\Psi^\ast\omega$ is a horizontal one-form
(i.e.,\ $\Psi^\ast\omega$
vanishes
along the fibres of the bundle $\Psi\colon P'\times G\to P'\times_{G'}
G$), which is invariant under the action of the structure group
$G'$. With these properties of $\Psi^\ast\omega$ it is now quite easy
to complete the proof.

The final step consists in the construction of local gauge potentials
for an invariant connection. As it turns out, using results from the
proof sketched above, this can easily be achieved. One finds

\corollary{1.}
{\em
Let $P$ be a $K$-symmetric principal bundle classified by $(\l,\tilde
Q)$, let $\omega$ be an invariant connection in $P$ classified by
$(\tilde\omega,\tilde\phi)$ and let
$
\tilde \sigma
$
and
$
\hat\sigma
$
be local sections of the bundles $\tilde Q$ and $K(K/H,H)$,
respectively. Then there is a local section $\sigma$ of $P$ such that
the gauge potential $A=\sigma^\ast\omega$ is given by
\[
A=\tilde\sigma^\ast\tilde\omega + (\tilde\sigma^\ast \tilde\phi+
L\l)\circ (\hat\sigma^{-1}\,d\hat\sigma)
\;.
\]
}

The presented theorems, together with this corollary, reduce the
construction of symmetric gauge potentials to well studied, purely
group theoretical problems. However, we have assumed that the symmetry
group acts by bundle automorphisms. Physically, it would be more
natural
to require that the action is realised only projectively, whereby the
projective factors are global gauge transformations. Correspondingly,
one should then also weaken the invariance condition for the
connection.

\end{document}